\documentclass[aps,prl,twocolumn,showpacs,groupedaddress]{revtex4}

\usepackage{graphicx}
\usepackage{dcolumn}
\usepackage{bm}

\begin{document}



\title{Mott-insulator to commensurate-solid transition in a $^4$He layer 
on $\alpha$-graphyne: Pseudo-spin symmetry breaking under a particle-induced pseudo-magnetic field
} 
\author{Yongkyung Kwon} 
\email{ykwon@konkuk.ac.kr}
\author{Hyeondeok Shin}
\author{Hoonkyung Lee}
\affiliation{
Division of Quantum Phases and Devices, School of Physics,
Konkuk University, Seoul 143-701, Korea
}%

\date{\today}


\begin{abstract}
{Path-integral Monte Carlo calculations were performed to
study the adsorption of $^4$He atoms on $\alpha$-graphyne. 
We find that one $^4$He atom can be embedded onto the in-plane center of each hexagon of the graphyne.
In the first $^4$He layer above the $^4$He-embedded graphyne surface,
a Mott insulating state was observed at the areal density of 0.0706~\AA$^{-2}$ 
with three $^4$He atoms occupying each hexagonal cell
while the helium atoms form a commensurate triangular solid at a density of 0.0941~\AA$^{-2}$.
Here we show that the Ising pseudo-spin symmetry introduced for two degenerate
configurations of three $^4$He atoms in a hexagonal cell can be broken
by additional $^4$He atoms placed at the hexagon vertices
and the Mott-insulator to commensurate-solid transition is 
a transition from a nonmagnetic spin liquid of frustrated antiferromagnets to a spin-aligned ferromagnet
under a particle-induced pseudo-magnetic field.
}
\end{abstract}


\maketitle



Graphite is known to be a strong substrate for $^4$He, on which
multiple distinct two-dimensional helium layers were observed~\cite{zimmerli92}.
The interplay between $^4$He-$^4$He interactions and $^4$He-substrate interactions
results in rich structural phase diagrams for the helium adlayers on graphite, such as
two-dimensional fluids, commensurate and incommensurate solids.
Using torsional oscillator experiments, Crowell and Reppy first observed  
finite superfluid fractions in the second $^4$He layer,
from which they once speculated possible existence of a two-dimensional supersolid phase~\cite{crowell96}.
In recent times, various forms of carbon allotropes have been discovered,  
among which graphene, carbon nanotubes, and fullerene molecules have attracted a great deal of attention
as novel nanomaterials because of their distinct electronic and mechanical properties.
Recent theoretical studies~\cite{gordillo09,kwon12,gordillo12,happacher13} show that
multiple distinct $^4$He layers are formed on a single graphene sheet
and the helium monolayer on graphene exhibits a quantum phase diagram similar to the one observed 
in the corresponding layer on graphite, including the commensurate-incommensurate solid transition.   
While $^4$He atoms confined in nanotube bundles showed characteristics of 
a quasi-one-dimensional system~\cite{cole00,gordillo08},
near-spherical helium layers on the outer surfaces of fullerene molecules
were found to exhibit different quantum states depending on the number of $^4$He adatoms~\cite{kwon10,shin12b,calvo12,leidlmair12}.
In particular, the helium monolayer on a C$_{20}$ molecular surface 
was predicted to be completed with 32 $^4$He atoms and to show nanoscale supersolidity
induced by mobile vacancy states near its completion~\cite{kwon10}.

Graphynes are two-dimensional networks of $sp$- and $sp^2$-bonded carbon atoms
that possess some intriguing electronic properties. 
In particular, highly-asymmetric Dirac cones predicted for some graphynes
would cause electrons to conduct only in a preferred direction, 
leading to electron collimation transport without any external field~\cite{malko12}.
Graphyne also has a very large surface area, with its hexagon being much larger than that of graphene,
which could allow various potential applications as new energy materials,
including as a Li-ion battery anode~\cite{zhang11,hwang13} and high-capacity hydrogen storage~\cite{hwang12,koo13}.
Because of its large hexagon size,  
a single sheet of graphyne, unlike a graphene sheet, is predicted to be permeable to $^4$He atoms.
This could result in more complex phase diagrams for the $^4$He adlayers on graphyne
than those of the corresponding helium layers on graphene or graphite.

Here we performed path-integral Monte Carlo (PIMC) calculations 
to study $^4$He adsorption on $\alpha-$graphyne, 
a honeycomb structure with a hexagon side consisting of one $sp^2$-bonded carbon atom 
and two $sp$-bonded carbon atoms~\cite{coluci04}.
Unlike the case of graphene, 
in-plane adsorption of $^4$He atoms is observed on the graphyne surface
with a single $^4$He atom being embedded to the center of each hexagon.
The first layer of $^4$He atoms adsorbed on the $^4$He-embedded graphyne surface
exhibits various quantum phases depending on the helium coverage;
this helium layer is in a Mott insulating state at the areal density of 0.0706~\AA$^{-2}$ 
with each hexagonal unit cell accommodating three $^4$He atoms
and the helium atoms form a commensurate triangular solid at 0.0941~\AA$^{-2}$.  

In this study, the helium-graphyne interaction is described by  
a sum of pair potentials between the carbon atoms and a $^4$He atom.
For the $^4$He-C interatomic pair potential, we use an isotropic six-twelve Lennard-Jones potential, 
which was proposed by Carlos and Cole to fit helium scattering data from graphite surfaces~\cite{carlos80,cole83}.
This empirical pair potential has been widely used to study helium adsorptions on various carbon isomers.
A well-known Aziz potential~\cite{aziz92} is used  for the $^4$He-$^4$He interaction. 
In the path-integral representation, the thermal density matrix at a low temperature $T$ 
is expressed as a convolution of $M$ high-temperature density matrices with 
an imaginary time step $\tau = (M k_B T)^{-1}$.  
Both $^4$He-$^4$He and $^4$He-C potentials are used to compute
the exact two-body density matrices~\cite{ceperley95,zillich05} at the high temperature $MT$,
which was found to provide an accurate description of 
the $^4$He-substrate interaction as well as the $^4$He-$^4$He interaction
with a time step of $\tau^{-1}/k_B = 40$ K.
Periodic boundary conditions with a fixed rectangular simulation cell are used to minimize finite size effects

\begin{figure}
\includegraphics[width=3.2in, height=4.0in]{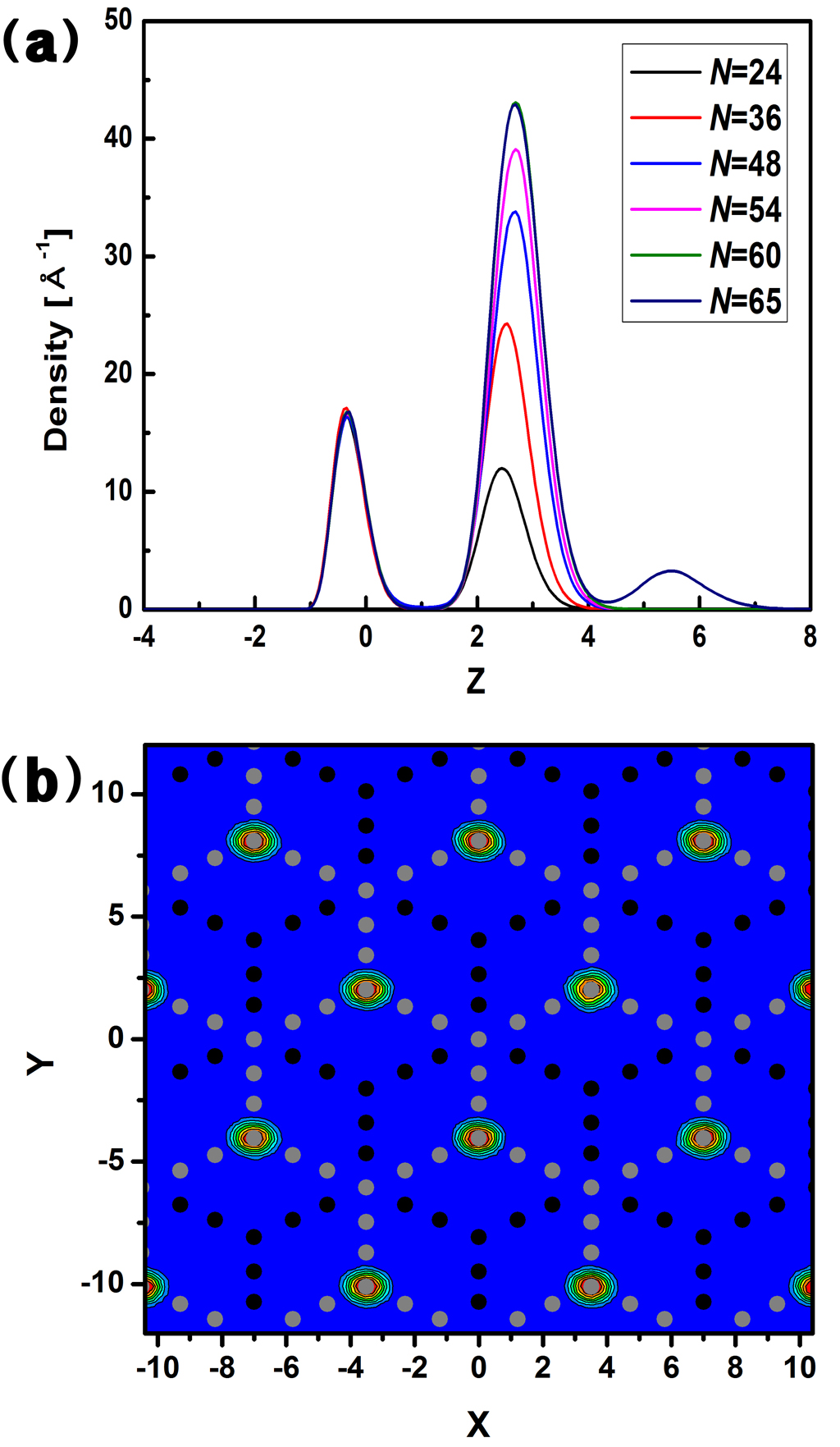}
\caption{(Color online) (a) One-dimensional density of $^4$He atoms adsorbed on a bilayer $\alpha$-graphyne
as a function of the distance $z$ (in \AA) from the graphyne surface and (b) two-dimensional density distribution
of $^4$He atoms constituting the first density peak located at $z \sim -0.3$~\AA~(red: high density, blue: low density).  
Here $N$ represents the number of $^4$He adatoms per $3 \times 2$ rectangular simulation cell 
with dimensions of $21.01 \times 24.26$~\AA$^2$ and the computations were done at a temperature of $0.5$~K. 
The black and brown dots in (b) represent the two-dimensional positions of the carbon atoms 
of the upper and the lower graphyne layer, respectively. 
} 
\label{fig:1Dden}
\end{figure}

With the $^4$He-substrate potential described above,
we performed PIMC calculations for $^4$He atoms adsorbed on an $AB$-stacked bilayer $\alpha-$graphyne,
whose interlayer spacing was set to be the same as that of graphite.
Figure~\ref{fig:1Dden}(a) shows the one-dimensional $^4$He density distributions as a
function of the distance $z$ from the graphyne surface,
for different numbers of $^4$He adatoms per $3 \times 2$ rectangular simulation cell, 
which has dimensions of $21.01 \times 24.26$~\AA$^2$.
Here the upper graphyne layer is positioned at $z=0$ and the lower layer at $z=-3.35$~\AA.
One can observe that increase in the number of $^4$He adatoms leads to the development 
of distinct layered structures, similar to those observed on a graphite or graphene surface.
The sharp peak located at $z \sim -0.3$~\AA~is completed with 12 $^4$He atoms. 
This corresponds to the areal density of 0.0235~\AA$^{-2}$, or to one helium atom per the hexagonal cell,
which leads us to a conjecture that each hexagonal cell on the graphyne surface 
can accommodate one $^4$He atom at its in-plane center.
This has been confirmed by the two-dimensional density distribution of 12 $^4$He atoms in Fig.~\ref{fig:1Dden}(b),
which shows clear density peaks located at the centers of the hexagons.
The black and brown dots of the honeycomb structures represent 
the two-dimensional positions of the carbon atoms in the upper and the lower graphyne layer, respectively.
The layer of $^4$He atoms embedded onto the hexagon centers of the graphyne surface is called 
the zeroth helium layer, 
which is located slightly below the upper graphyne layer 
because of the attractive interaction with the lower graphyne layer.
With the growth of the next helium layer (the first layer) located at $z \sim 2.7$~\AA,
there was little change in the zeroth layer except for a slight tightening 
due to the compression from the first-layer $^4$He atoms.

\begin{figure*}
\includegraphics[width=6.5in, height=3.2in]{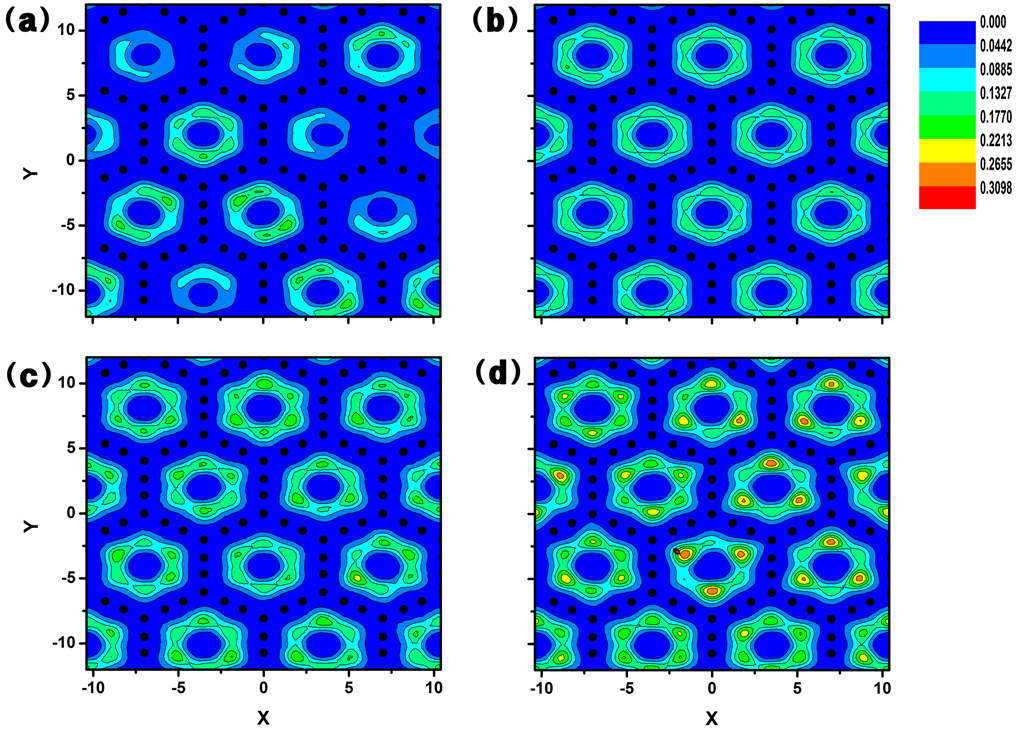}
\caption{(Color online) Two-dimensional density distributions of the first-layer $^4$He atoms
adsorbed on the $^4$He-embedded $\alpha$-graphene surface at areal densities of 
(a) 0.0235, (b) 0.0471, (c) 0.0588, and (d) 0.0706~\AA$^{-2}$.
The black dots represent the
positions of the carbon atoms in the upper graphyne layer. The length unit is \AA~and
all contour plots are in the same color scale denoted by the color table in the upper right hand corner.
} 
\label{fig:2Dden}
\end{figure*}

We now analyze the structure of the first $^4$He layer 
above the $^4$He-embedded graphyne surface.
Two-dimensional density distributions of the first $^4$He layer, at a temperature
of $T=0.5$~K, are shown in Fig.~\ref{fig:2Dden}.
We note that an adsorption site for $^4$He (or a hexagonal unit cell) determined by the substrate potential 
is large enough to accommodate multiple $^4$He atoms.
In fact, each hexagonal cell is found to  be able to accommodate up to 3 $^4$He atoms. 
At the first-layer areal density of $\sigma_1=0.0235$~\AA$^{-2}$, which 
corresponds to one $^4$He atom per hexagonal cell,
the average helium occupancies of some hexagonal cells are higher than one while 
the others have the helium occupancies lower than one (see Fig.~\ref{fig:2Dden}(a)).
This irregularity can be understood by the fact that attractive on-site interaction
between two $^4$He atoms accommodated by a single hexagonal cell, 
along with the substrate-potential barrier, results in low probability
for a $^4$He atom to hop to a neighboring cell. 
When the helium density is doubled to $\sigma_1=0.0471$~\AA$^{-2}$,
every hexagonal cell seen in Fig.~\ref{fig:2Dden}(b) contains two $^4$He atoms.
As the helium coverage increases further, shown in Fig.~\ref{fig:2Dden}(c), 
all hexagonal cells show fractional occupancies close to $\sim 2.5$,
which suggests frequent hopping of the third $^4$He atom in a cell to its neighbor.
Figure~\ref{fig:2Dden}(d) shows that   
each hexagon involves the maximum number of $^4$He atoms, 
namely 3 $^4$He atoms, at the areal density of 0.0706~\AA$^{-2}$. 
This is concluded to be a Mott insulating state~\cite{fisher89}
where a strong repulsive on-site interaction prohibits a $^4$He atom 
from hopping from one adsorption site (a hexagonal cell) 
to the neighboring ones. 
The winding number estimator of Ref.~\cite{ceperley95} was used 
to compute the superfluidity of the $^4$He layer on the $^4$He-embedded graphyne surface,
from which significant superfluid fractions were observed only at the areal densities corresponding to
fractional $^4$He occupancies as in the case of Fig.~\ref{fig:2Dden}(c). 
In particular, the superfluidity was found to be completely quenched
at the Mott insulating density of $\sigma_1=0.0706$~\AA$^{-2}$, where
the $^4$He layer undergoes a superfluid to Mott-insulator transition. 

\begin{figure}
\includegraphics[width=3.2in]{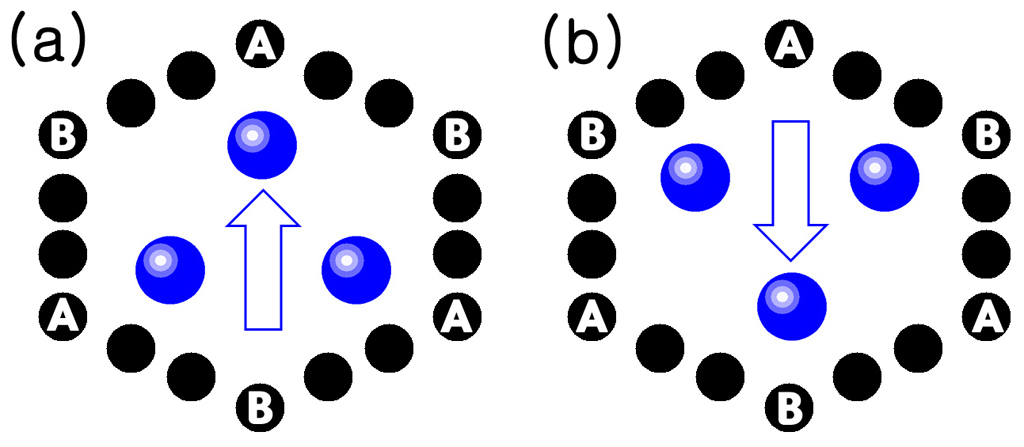}
\caption{(Color online) Ising pseudo-spin states ((a) spin up and (b) spin down)
assigned to two degenerate ground-state configurations for three first-layer $^4$He atoms
accommodated in a hexagonal unit cell on the $\alpha$-graphyne surface.
The black and the blue dots correspond to the positions of the carbon atoms and the $^4$He atoms, respectively.
The characters A and B represent two different triangular sublattice sites 
of the honeycomb structure of $\alpha$-graphyne.
}
\label{fig:pseudo}
\end{figure}

As shown in Fig.~\ref{fig:pseudo}, there are two degenerate ground-state configurations for 3 first-layer $^4$He atoms 
in a hexagonal unit cell on the graphyne surface. This allows us to introduce Ising pseudo-spins; 
(a) spin up for one configuration and (b) spin down for the other.
It is found that the inter-cell interaction between $^4$He atoms in two neighboring cells 
energetically favors anti-aligned pseudo-spin configurations 
and results in a weak antiferromagnetic interaction between the neighboring spins.
The helium density distribution for the Mott insulating state, shown in Fig.~\ref{fig:2Dden}(d),
shows that each pseudo-spin fluctuates between the up- and the down-state
with a tendency of being anti-aligned with its neighbors.
These fluctuations
are understood to be due to the geometrical frustration at the triangular lattice.
Another PIMC simulation starting from an initial spin-aligned configuration 
also produced a $^4$He density distribution 
consistent with the pseudo-spin state of Fig.~\ref{fig:2Dden}(d)
which was obtained from a random initial $^4$He configuration.
This suggests that 
the Mott insulating state corresponds to a nonmagnetic spin liquid of frustrated antiferromagnets~\cite{balents10}
in terms of the pseudo-spin degrees of freedom.

\begin{figure}
\includegraphics[width=3.2in, height=4.0in]{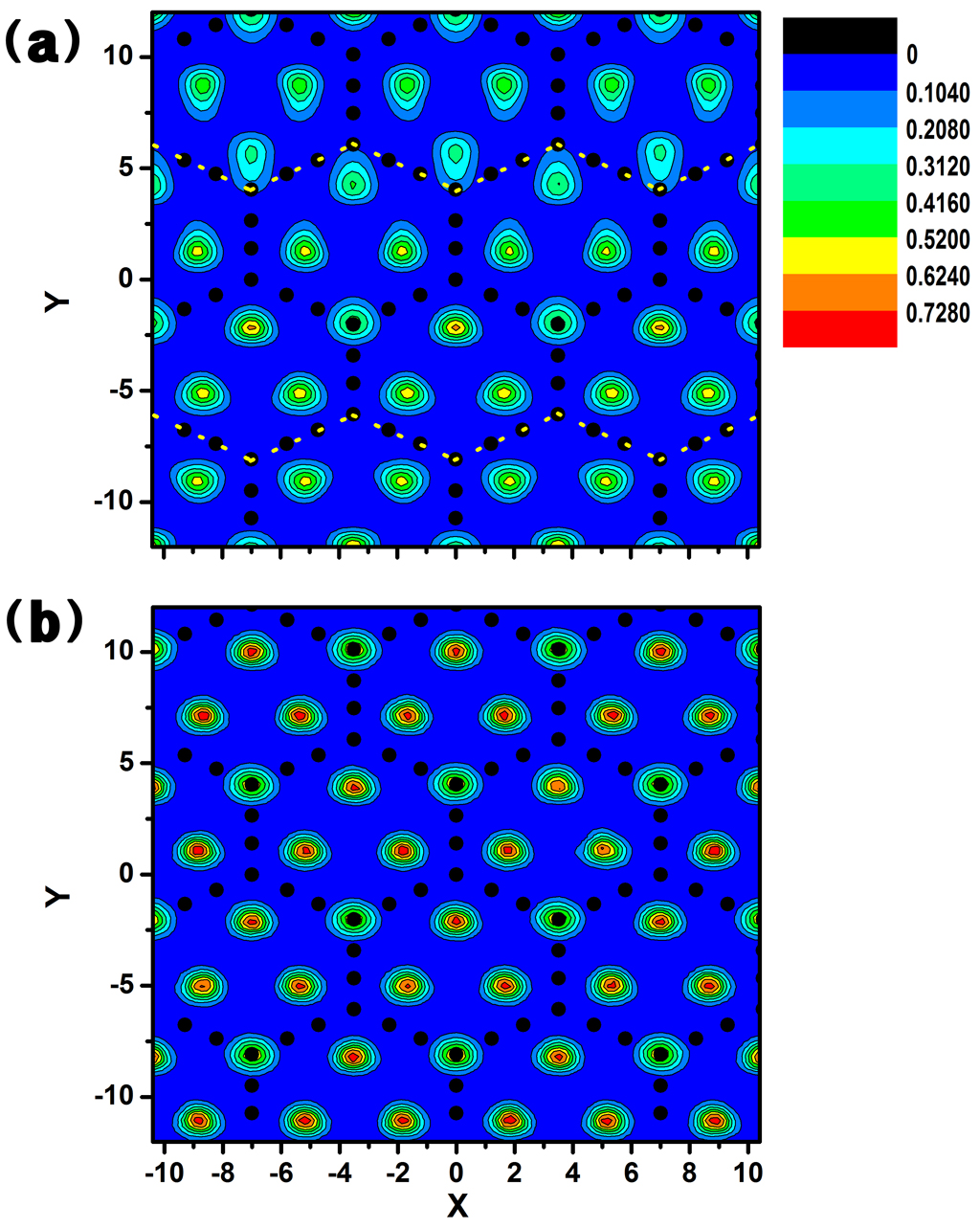}
\caption{(Color online) Two-dimensional density distributions of the first-layer $^4$He atoms
adsorbed on the $^4$He-embedded graphyne surface at high areal densities of (a) 0.0824~\AA$^{-2}$
and (b) 0.0941~\AA$^{-2}$. 
The black dots represent the locations of carbon atoms in the upper graphyne layer. The length unit is \AA~and
both contour plots are in the same color scale denoted by the color table in the upper right hand corner.
The yellow dotted lines in (a) separate two distinct ferromagnetic domains. 
} 
\label{fig:high_den}
\end{figure}

We now discuss the high-density structures of the first helium layer 
above the $^4$He-embedded $\alpha$-graphyne.
Figure~\ref{fig:high_den} shows the two-dimensional density distributions
of the first-layer $^4$He atoms at two high areal densities. 
When the areal density increases beyond
the Mott-insulating density of $\sigma_1=0.0706$~\AA$^{-2}$,
the additional helium atoms are found to be placed at the vertices of the hexagons, 
that is, above the $sp^2$-bonded carbon atoms.
In Fig.~\ref{fig:high_den}(a), one can see 6 $^4$He atoms
located at the hexagon vertices.
The presence of a $^4$He atom at a hexagon vertex 
affects the pseudo-spin states of the surrounding hexagonal cells.
There are two different triangular sublattice points in the honeycomb structure of $\alpha$-graphyne, 
which are denoted by $A$ and $B$ in Fig.~\ref{fig:pseudo}. 
The additional helium atoms placed at $A$ sites favor
the surrounding three pseudo-spins to be in the spin-up state to minimize the $^4$He-$^4$He interaction,
while those at $B$ sites prefer the spin-down state in the neighboring cells.
Hence the $^4$He atoms at the $A$ and $B$ sites play the role of a local pseudo-magnetic field 
that forces the surrounding spins to be aligned in a certain direction.
As the helium coverage increases beyond the Mott insulating density,
we observed the formation of 
some local ordering of the pseudo-spins due to some additional helium atoms occupying the vertex carbon sites.
In Fig.~\ref{fig:high_den}(a) for the areal density of 0.0824~\AA$^{-2}$,
two well-defined ferromagnetic domains with each of them consisting of the pseudo-spins aligned in the same direction,
are seen to be separated by the yellow dotted lines.
A homogeneous ferromagnetic phase is observed at the areal density of 0.0941~\AA$^{-2}$.
In Fig.~\ref{fig:high_den}(b), every $A$ site, one of two triangular sublattices of the honeycomb structure, 
is occupied by a single $^4$He atom while every $B$ site, the other sublattice, is vacant without any $^4$He atom.
This sublattice symmetry breaking causes all pseudo-spins to be aligned in the same direction. 
One can also see in Fig.~\ref{fig:high_den}(b) that the first-layer $^4$He atoms in the spin-aligned ferromagnetic phase constitute a triangular solid 
whose crystalline structure is commensurate with the underlying honeycomb structure of the graphyne surface. 
We note that all lattice points of this triangular solid are not the adsorption sites predetermined by the substrate potential
and that the interaction potential between a first-layer $^4$He atom and the $^4$He-embedded graphyne surface
has the highest value at the hexagon vertices. From this we understand that 
the $^4$He-$^4$He interaction as well as the $^4$He-substrate interaction is crucial in the manifestation of 
the triangular crystalline structure shown in Fig.~\ref{fig:high_den}(b).
We observe that $^4$He atoms are promoted to the second layer when the first-layer helium coverage 
increases beyond $\sigma_1=0.0941$~\AA$^{-2}$. 
 
Our PIMC calculations reveal that the in-plane $^4$He adsorption takes place on $\alpha$-graphyne
and a single helium atom is embedded to its in-plane hexagon center.
The first layer of $^4$He atoms adsorbed on the $^4$He-embedded graphyne surface is found to exhibit
various quantum phases depending on the helium coverage.
The Mott insulating state where each hexagonal unit cell accommodates three $^4$He atoms
is considered to be a spin liquid of frustrated antiferromagnets at a triangular lattice 
in terms of the Ising pseudo-spins assigned to two degenerate helium configurations.
The pseudo-spin symmetry can be broken by the presence of additional $^4$He atoms occupying the hexagon vertices,
which causes a ferromagnetic domain to be developed at higher helium densities.
The pseudo-magnetic field induced by the additional particles is similar 
to the strain-induced pseudo-magnetic field proposed recently 
to engineer graphene electronic structures~\cite{guinea10,levy10},
in a sense that it breaks the sublattice symmetry of a honeycomb structure.
At the helium coverage of $\sigma_1=0.0941$~\AA$^{-2}$, 
the first $^4$He layer is in a ferromagnetic phase with
all Ising pseudo-spins being aligned in the same direction under a particle-induced pseudo-magnetic field, 
where the $^4$He adatoms are found to constitute a commensurate triangular solid. 
The formation of vacancy defects in this two-dimensional solid 
and their possible contribution to superfluidity is now under investigation.

\begin{acknowledgments}
Y. Kwon thanks Hunpyo Lee for helpful discussion.
This work was supported by the Basic Science Research Program (2012R1A1A 2012006887)
and the WCU Program (R31-2008-000-10057-0) through the National Research Foundation
of Korea funded by the Ministry of Education, Science and Technology.
\end{acknowledgments}


\end{document}